\documentclass[letterpaper,titlepage,11pt]{article}
\usepackage{latexsym,amssymb,amstext,amsmath}
\usepackage{slashed}
\usepackage{mathrsfs}
\usepackage{color}
\usepackage{enumerate}
\usepackage{graphicx}
\usepackage{tikz}
\usepackage{etex}
\usepackage{latexsym}
\usepackage{cite}

\allowdisplaybreaks

\usepackage[
colorlinks=true,
linkcolor=blue,
urlcolor=red,
filecolor=green,
citecolor=red,
pdfstartview=FitV,
pdftitle={},
pdfauthor={Mehmet Ozkan},
pdfsubject={},
pdfkeywords={},
pdfpagemode=None,
bookmarksopen=true
]{hyperref}

\newcommand{\bea}{\setlength\arraycolsep{2pt} \begin{eqnarray}}
	\newcommand{\eea}{\end{eqnarray}}
\newcommand{\nn}{\nonumber}

\usepackage{hyperref}

\newsavebox{\uuunit}
\sbox{\uuunit}
{\setlength{\unitlength}{0.825em}
	\begin{picture}(0.6,0.7)
		\thinlines
		\put(0,0){\line(1,0){0.5}}
		\put(0.15,0){\line(0,1){0.7}}
		\put(0.35,0){\line(0,1){0.8}}
		\multiput(0.3,0.8)(-0.04,-0.02){12}{\rule{0.5pt}{0.5pt}}
		\end {picture}}

	\def\be{\begin{equation}}
		\def\ee{\end{equation}}
	\def\ba{\begin{array}}
		\def\ea{\end{array}}
	\def\bea{\begin{eqnarray}}
		\def\eea{\end{eqnarray}}
	\def\bd{\begin{displaymath}}
		\def\ed{\end{displaymath}}
	
	\def\nn{\nonumber}
	
	\def\a{\alpha}
	
	\def\g{\gamma}

	\def\e{\epsilon}

	\def\l{\lambda}
	\def\L{\Lambda}
	\def\m{\mu}
	\def\n{\nu}
	\def\r{\rho}
	\def\s{\sigma}
	
	\def\t{\tau}

	\def\o{\omega}
	\def\O{\Omega}

	\def\nn{\nonumber}

	\def\cL{\mathcal{L}}

	\makeatletter
	\@addtoreset{equation}{section}
	\makeatother
	
\begin{document}
		\begin{titlepage}
			
			\bigskip
			\begin{center}
				{\LARGE\bfseries  Chern--Simons-like formulation of exotic massive 3D gravity models}
				\\[10mm]
				\textbf{B\"u\c{s}ra Dedeo\u{g}lu$^1$, Mehmet Ozkan$^2$ and \"Ozg\"ur Sar{\i}o\u{g}lu$^1$}\\[5mm]
				\vskip 25pt
				
				{\em  \hskip -.1truecm $^1$ Department of Physics, Faculty of Arts and Sciences,
					Middle East Technical University, 06800, Ankara, T\"urkiye  \vskip 10pt }
				
				{\em  \hskip -.1truecm $^2$Department of Physics, Istanbul Technical University,  \\
					Maslak 34469 Istanbul, T\"urkiye  \vskip 10pt }

				{email: {\tt busrad@metu.edu.tr, ozkanmehm@itu.edu.tr, sarioglu@metu.edu.tr}}
				
			\end{center}
			
			\vspace{3ex}

			\begin{center}
				{\bfseries Abstract}
			\end{center}
			\begin{quotation} \noindent
				
						We investigate the Chern--Simons-like formulation of exotic general massive gravity models within the framework of third-way to three-dimensional gravity. We classify our construction into two main approaches: one using torsional cosmological Einstein and exotic massive gravity equations, and the other a torsion-free approach. The former approach, while mathematically appealing, precludes the construction of critical exotic models where the central charges vanish. In contrast, the latter approach has a wider parameter space and allows for the construction of critical models. An explicit example of an exotic general model is provided to illustrate both methods. Our methodology represents the first step towards establishing the most general Chern--Simons-like formulation of third-way to three-dimensional gravity, which would enable the study of identifying its bulk/boundary unitary sector.

			\end{quotation}
			
			\vfill
			
			\flushleft{\today}
		\end{titlepage}
		
		\tableofcontents
		
		\newpage

		\section{Introduction}
		\paragraph{}
		Three-dimensional gravity models serve as a rich testing ground to explore various ideas related to black holes and AdS/CFT duality. On one hand, Einstein's gravity with a negative cosmological constant lacks propagating local degrees of freedom, yet it still has black hole solutions \cite{Banados:1992wn,Banados:1992gq} and boundary degrees of freedom \cite{Brown:1986nw}. On the other hand, when local dynamics is introduced by means of higher-derivative extensions, arbitrary curvature combinations give rise to ghost degrees of freedom. This issue can be resolved by formulating Einstein's gravity with a negative cosmological constant as a Chern--Simons gauge theory \cite{Achucarro:1986uwr,Witten:1988hc}, and its higher-order extensions in a Chern--Simons-like formulation \cite{Hohm:2012vh,Bergshoeff:2014bia,Afshar:2014ffa,Merbis:2014vja}. These theories typically include a dreibein, a (dual) spin-connection as well as $N$-number of Lorentz-vector one-form auxiliary fields. Here, the number $N$ also characterizes the number of derivatives acting on the metric beyond two-derivatives. Hence, the increasing number of auxiliary fields indicate the increasing number of derivatives in the corresponding gravitational theory. 
		
		While the Chern--Simons-like formulation successfully addresses the propagation of the additional ghost degrees of freedom, a significant challenge remains in identifying the healthy sector within three-dimensional higher-derivative gravity models. This challenge, referred to as the bulk-boundary unitarity \cite{Li:2008dq}, arises from the fact that unitarity in AdS$_3$ space corresponds to the positivity of central charges in the asymptotic Virasoro $\oplus$ Virasoro algebra implied by the Brown-Henneaux conditions. Therefore, a theory achieving bulk-boundary unitarity must adhere to the no-tachyon condition and ensure the positivity of the central charges. Unfortunately, the Chern--Simons-like formulation alone does not provide a solution to this problem, and gravitational models that are represented by a single metric Lagrangian are known to be inadequate in satisfying the bulk-boundary unitarity on general grounds.
		
		The resolution to both the propagating ghost degrees of freedom and the bulk-boundary clash arises from a rather radical change of perspective, known as the third-way to three-dimensional gravity \cite{Bergshoeff:2014pca,Bergshoeff:2015zga}. For a gravitational theory, the metric field equation $E_{\mu\nu} = 0$ satisfies a Bianchi-type identity, $\nabla^\mu E_{\mu\nu} = 0$ as a consequence of diffeomorphism invariance of the action. This identity is off-shell in the sense that it does not refer to the field equations. Conversely, in the presence of a matter sector, the metric field equation also includes an energy-momentum tensor $E_{\mu\nu} = T_{\mu\nu}$. In this case, the divergence of the left-hand side of the metric field equation vanishes off-shell but its right-hand side vanishes on-shell, that is, once the matter field equations are imposed. The third-way to three-dimensional gravity mixes these two perspectives, and declares that the metric field equation for a purely gravitational theory vanishes once the metric field equation itself is imposed, i.e., $\nabla^\mu E_{\mu\nu} \propto E_{\mu\nu} \approx 0$. The generic form of such an equation is schematically given as \cite{Ozkan:2018cxj}
		\begin{eqnarray}
			E_{\mu\nu} \equiv \L_0 g_{\m\n} + G_{\m\n} + \mathcal{G}_{\mu\nu} +\mathcal{H}_{\mu\nu} + \mathcal{L}_{\mu\nu} = 0 \,,
			\label{3rdWayEOM}
		\end{eqnarray}
		where $G_{\mu\nu}$ is the Einstein tensor, and $\mathcal{H}_{\m\n}$ and $\mathcal{L}_{\m\n}$ are defined as
		\begin{align}
			\mathcal{H}_{\mu\nu} &= \e_\m{}^{\r\s} \nabla_\r \mathcal{S}_{\n\s} \,, & \mathcal{L}_{\m\n} & = \frac12 \e_\m{}^{\r\s} \e_\n{}^{\l\t} \mathcal{S}_{\r\l} \mathcal{S}_{\s\t} \,.
		\end{align}
		Here, $\mathcal{S}_{\m\n}$ is a Schouten-like tensor that is obtained from an Einstein-like tensor $\mathcal{G}_{\m\n}$ with $\nabla^\m \mathcal{G}_{\m\n} = 0$ as
		\begin{eqnarray}
			\mathcal{S}_{\m\n} = \mathcal{G}_{\m\n} - \frac12 g_{\m\n} g^{\r\l} \mathcal{G}_{\r\l} \,. 
		\end{eqnarray}
		Unlike the field equations that are derived from a diffeomorphism-invariant action with a single metric $g_{\m\n}$, the divergence of the field equation \eqref{3rdWayEOM} does not vanish identically, but is proportional to the field equation itself
		\begin{equation}
			\nabla^\m E_{\m\n} =  - \e_{\nu}{}^{\r\s} \mathcal S_{\r}{}^\l \left( G_{\s\l} + \mathcal{H}_{\s\l} \right) = - \e_{\nu}{}^{\r\s} \mathcal S_{\r}{}^\l \left( G_{\s\l} +  \mathcal{G}_{\s \l} + \mathcal{H}_{\s\l} + \L_0 g_{\s\l} + \mathcal{L}_{\s\l}  \right) \approx 0 \,.
   \label{Div3rdWay}
		\end{equation}
		Here, we add $\e_{\nu}{}^{\r\s} \mathcal S_{\r}{}^\l \left( \L_0 g_{\s\l} + \mathcal{G}_{\s\l} + \mathcal{L}_{\s\l} \right)$ to complete the parentheses into the field equation $E_{\m\n}$. This step is harmless as the additional terms vanish identically due to the symmetry property of the Schouten-like tensor $\mathcal{S}_{\m\n} = \mathcal{S}_{\n\m}$. The price to pay for this construction is that these models cannot arise from the variation of a diffeomorphism invariant Lagrangian of a single-metric theory. Nevertheless, they can nicely be incorporated into a Chern--Simons-like formulation which enables one to work out the bulk-boundary clash as well as the Hamiltonian formulation for the degrees of freedom analysis. The known examples of the third-way models include the Minimal Massive Gravity (MMG) \cite{Bergshoeff:2014pca}, Exotic General Massive Gravity (EGMG) \cite{Ozkan:2018cxj}, Bachian gravity \cite{Alkac:2018eck}, and further extensions are shown to be obtained from a truncation procedure in the Chern--Simons-like formulation \cite{Afshar:2019npk}. 
		
		Although we have the field equations for third-way models in a general form, as given by \eqref{3rdWayEOM}, a generic Chern--Simons-like formulation for these models is still missing. This represents a notable gap in the existing literature. Currently, MMG stands out as the sole model capable of resolving the bulk-boundary clash without introducing ghost degrees of freedom\footnote{There are bimetric models that are unitary and have no unphysical degrees of freedom \cite{Bergshoeff:2013xma, Afshar:2014dta, Ozkan:2019iga}. However, here we focus on single-metric theories and exclude the multi-metric option throughout this paper.}. The absence of a generic Chern--Simons-like formulation for third-way models hinders our ability to conduct a detailed analysis and to identify the set of healthy models. In this paper, we aim to initiate a systematic study by establishing the Chern--Simons-like formulation of the so-called exotic models of 3D gravity. The reason for this focus is that these models cannot be formulated by a metric alone due to their parity property, and are third-way by nature. Thus, it is a natural starting point to study the systematic formulation of third-way to 3D gravity.
		
	Our construction relies on two key elements: the recent observation that three-dimensional massive gravity models are closely related to a truncation of an infinite-dimensional algebra \cite{Bergshoeff:2021tbz} and the understanding that third-way models can be viewed as torsional Einstein gravity, where the torsion is given by a Schouten-like tensor \cite{Deger:2022gim}. Since this algebra and its properties are essential for constructing the third-way Chern--Simons-like models, we briefly discuss the algebra and its gauging, demonstrating how to incorporate the field equations into a Chern--Simons-like Lagrangian in Section \ref{S2}. In Section \ref{S3}, we introduce a torsional modification to the group-theoretical curvatures of the infinite-dimensional algebra described in \cite{Bergshoeff:2021tbz}. This modification allows us to present the exotic and exotic general massive 3D gravity models, which constitute one of the main results of this paper. 

 While the torsional case is mathematically appealing due to the simple torsion interpretation of the third-way equation of motion, its parameter space is more restricted compared to that of the torsion-free case. In particular, the torsional construction fixes the coefficient of a specific term, the Lorentz-Chern-Simons term, in terms of other free parameters of the theory if we demand AdS to be a maximally symmetric solution to the field equations. On the other hand, this particular coefficient is proportional to the central charges of the dual CFT in the case of exotic models. Consequently, the torsional models do not allow setting that particular coefficient to zero, implying that the critical models, where the central charges vanish, are excluded in the torsional construction. We demonstrate this difference explicitly for the leading-order model. We give our comments and conclusions in Section \ref{Conclusions}.

		\section{Chern--Simons-like formulation of massive 3D gravity models} \label{S2}
		\paragraph{}
		From the perspective of an infinite-dimensional algebra, we can elucidate the Chern--Simons-like formulation of three-dimensional massive gravity theories as follows. Consider the following infinite-dimensional algebra \cite{Bergshoeff:2021tbz}:
		\begin{align}
			[ J_a^{(m)}, J_b^{(n)}] & = \e_{abc} J^{c(m+n)}\,, & [ J_a^{(m)}, P_b^{(n)}] & = \e_{abc} P^{c(m+n)}\,,\nonumber\\
			[ P_a^{(m)}, P_b^{(n)}] & =  \e_{abc} J^{c(m+n-1)}\,,
			\label{Algebra}
		\end{align}
		Here, we define $P_a^{(0)} \equiv P_a$ as the generators of translations, and $J_a^{(0)} \equiv J_a$ as the generators of dual Lorentz transformations. The group-theoretical curvatures, which also represent the covariant field equations of the infinite-dimensional algebra upon setting to zero, can be adequately formulated by using a parameter $\lambda$ as \cite{Bergshoeff:2021tbz}
		\begin{eqnarray}
			0 &=& \left( de + \omega \times e \right) + \sum_{n\geq 1} \l^{2n} \left( D(\o)h_{(n)} + e \times f_{(n)}\right) + \sum_{m,n \geq 1} \l^{2(m+n)} h_{(m)} \times f_{(n)} \,,\nonumber\\
			0 &=& \left( R(\o) +  e \times h_{(1)}\right) + \frac12 \sum_{m,n \geq 1} \l^{2(m+1)}  f_{(m)} \times f_{(n)} \nonumber\\
			&& + \sum_{n \geq 1} \lambda^{2n} \left( D(\o)f_{(n)} + e \times h_{(n+1)} \right) + \frac12  \sum_{m,n \geq 1} \l^{2(m+n-1)} h_{(m)} \times h_{(n)} \,,
			\label{Hierachy-Lambda}
		\end{eqnarray}
		where $D(\omega)$ represents the Lorentz covariant derivative. Here, the generators and the gauge fields are associated as follows 
		\begin{eqnarray}
			A = e^a P_a + \o^a J_a + h^a_{(n)} P^{(n)}_a + f^a_{(n)} J^{(n)}_a  \,,
		\end{eqnarray}
		where $n\geq 1$. The gauge covariant field equations, or curvature constraints, are obtained by equating all terms that involve a specific order of $\lambda$  in \eqref{Hierachy-Lambda} to zero. When these field equations are imposed, they give rise to an infinite series of solvable equations that determine the associated gauge fields of $P_{a}^{(n)}$ (for $n \geq 1$) and $J_{a}^{(n)}$ (for $n\geq 0$) in terms of the dreibein, $e^a$. 
		
		If the infinite-dimensional algebra \eqref{Algebra} is truncated at a finite order, the Jacobi identities are no longer satisfied. If the truncated field equations are integrated to a Lagrangian, they describe the massive models of 3D gravity that are compatible with a holographic c-theorem \cite{Bergshoeff:2021tbz}. With the following parity assignment
		\begin{align}
			\text{Even}:&\,\, \{e^a, h^a_{(n)}\} \,, & \text{Odd}:& \,\, \{\omega^a, f^a_{(n)}\} \,, 
			\label{Parity}
		\end{align}
		the massive 3D gravity models can then be classified based on the parity characteristics of the Lagrangian.
		
		Parity-even models of massive 3D gravity are obtained by truncating the algebra with the same number of $P_{(n)}$ and $J_{(n)}$ generators. In that case, the field equations \eqref{Hierachy-Lambda} imply the following $n^{th}-$order field equations
		\begin{eqnarray}
			0 &=& T(\o)\,,\nn\\
			0 &=& D(\o) h_{(n)} + e \times f_{(n)} + \sum_{m=1}^{n-1} h_{(m)} \times f_{(n-m)} \,,\nn\\
			0 &=& R(\o) +  e \times h_{(1)} \,,\nn\\
			0 &=& D(\o) f_{(n)} +  e \times h_{(n+1)} + \frac12 \sum_{m=1}^{n-1} f_{(m)} \times f_{(n-m)}  + \frac{1}{2}   \sum_{m=1}^{n} h_{(m)} \times h_{(n-m+1)}   \,,
			\label{Eom}
		\end{eqnarray}
		where $T(\o) = de + \omega \times e$. The first and the third equations are the lowest ones that are integrated by the highest auxiliary fields, $f_{(N)}$ and $h_{(N)}$, respectively. Similarly, setting $n=N$ yields the highest field equations where the solvable one becomes the field equation for $\o$ while the dynamical one becomes the field equation for $e$. At this point, a brief discussion on the truncation of the field equations \eqref{Hierachy-Lambda} is in order. In the infinite-dimensional case, we have 
		\begin{eqnarray}
			0 &=& T(\o) + \sum_{\ell \geq 1} \l^{2\ell} \left(D(\o)h_{(\ell)} + e \times f_{(\ell)}\right) +  \sum_{m\geq 1}  \sum_{\ell\geq 1}  \l^{2m + 2\ell}  h_{(m)}  \times f_{(\ell)} \,.
			\label{InfiniteE}
		\end{eqnarray}
		This indicates that the lowest-order equation is $T(\o) = 0$. If we truncate this series at order $n$, meaning that we are collecting the terms with coefficient $\l^{2n}$, then the second term simply contributes by setting $\ell = n$. For the last term, we set $m+\ell = n$, which means that the lower bound for $m$ is still $m=1$ but the upper bound is $n-1$ since the lower bound on $\ell$ is $\ell = 1$. Hence, we obtain the following equations
		\begin{align}
			0 &= T(\o)\,, &
			0 &= D(\o) h_{(n)} + e \times f_{(n)} + \sum_{m=1}^{n-1} h_{(m)} \times f_{(n-m)} \,.
			\label{ExpT}
		\end{align}
		Note that for the highest-order term $n=N$, we obtain the field equation for the spin-connection. Next, we turn to the second equation in the infinite expansion \eqref{Hierachy-Lambda}. This equation indicates that at zeroth-order, we have
		\begin{eqnarray}
			0 &=& R(\o) +  e \times h_{(1)} \,,
		\end{eqnarray}
		while at $n^{th}$-order, we have
		\begin{eqnarray}
			0 &=& 	D(\o) f_{(n)}  +  e \times h_{(n+1)}  + \frac12 \sum_{m=1}^{n-1} f_{(m)} \times f_{(n-m)} + \frac{1}2  \sum_{m=1}^{n} h_{(m)} \times h_{(n-m+1)} \,,
		\end{eqnarray}
		where the upper bound on the index $m$ arises from the fact that $k = n - m$ and the lower bound for $k$ is $k=0$, hence the upper bound for $m$ is $m=n$. Note that for the highest field equation, which we obtain by setting $n=N$, the second term simply vanishes. 
		
		These field equations can be integrated to a Lagrangian. The lowest-order equations can be integrated by the highest-order fields $h_{(N)}$ and $f_{(N)}$. The other field equations can be integrated as follows. We first note that the second equation in \eqref{Eom} arises from the field equation for $f_{(N-n)}$ where $n$ is restricted as $1 \leq n \leq N-1$. Similarly, the last equation in \eqref{Eom}  is the field equation for $h_{(N-n)}$ with the same restriction on $n$. Consequently, the Lagrangian is given by
		\begin{eqnarray}
			\cL_{\rm{even}}^{N} &=& h_{(N)} \cdot R(\omega ) + f_{(N)} \cdot T(\omega)  + \sum_{n=1}^{N-1} f_{(N-n)} \cdot D(\omega) h_{(n)}\nn\\
			&& + \frac12 \sum_{n=1}^{N-1} e \cdot f_{(N-n)} \times f_{(n)}  + \frac12 \sum_{n=1}^{N-1} \, \sum_{m=1}^{n-1} f_{(N-n)} \cdot h_{(m)} \times f_{(n-m)}  \nn\\
			&&  + \frac12 \sum_{n=0}^{N-1} e \cdot h_{(N-n)} \times h_{(n+1)} + \frac16  \sum_{n=1}^{N-1} \, \sum_{m=1}^{n} h_{(m)} \cdot h_{(N-n)} \times h_{(n-m+1)} \,.
			\label{woEH}
		\end{eqnarray}
		All the $N^{th}$-order Lagrangians, including the Einstein-Hilbert term, can be added to in a sequence without spoiling the structure of field equations
		\begin{eqnarray}
			\cL &=& \s e \cdot R(\o) + \frac16 \L_0 \, e \cdot e \times e + \sum_{N} a_N \cL_{\rm{even}}^N \,,
		\end{eqnarray}
		where $\s = \pm 1$ and $a_N$ are constants with the mass dimension $-2N$.
		
		
		In the case of parity-odd models, we have one less $f$-field than $h$-fields. Specifically, at order $N$, the highest $f$-field is $f_{(N-1)}$ while the highest $h$-field is $h_{(N)}$. In this case, the truncation of the infinite series occurs as follow. First, consider the equation \eqref{InfiniteE}. The truncation to $n^{th}$-order yields the same result as \eqref{ExpT}, however, we should keep in mind that for $n=N$, the second term in the second equation drops out. Consequently, we have
		\begin{align}
			0 &= T(\o)\,, &
			0 &= D(\o) h_{(n)} + e \times f_{(n)} + \sum_{m=1}^{n-1} h_{(m)} \times f_{(n-m)} \,,
			\label{Odd1}
		\end{align}
		as before.  Here, the lowest-order equation is the field equation for $h_{(N)}$ while the highest one, $n=N$, is the field equation for the dreibein $e^a$. Note that for $n=N$, the $e \times f_{(N)}$ term simply drops out. For the second equation in \eqref{Hierachy-Lambda}, we first note that as $h_{(n)}$ is truncated at order $\mathcal{O}(\l^{2n})$, $f_{(n)}$ needs to be truncated at order $\mathcal{O}(\l^{2n-2})$. Thus, we obtain
		\begin{eqnarray}
			0 &=& R(\o) +  e \times h_{(1)} \,,\nn\\ 
			0 &=& 	D(\o) f_{(n-1)}  +  e \times h_{(n)}  + \frac12 \sum_{m=1}^{n-2} f_{(m)} \times f_{(n-m-1)} + \frac12  \sum_{m=1}^{n-1} h_{(m)} \times h_{(n-m)} \,,
			\label{Odd2}
		\end{eqnarray}
		where the highest value of $n$ is restricted to $N-1$. Once again, the lowest equation is the equation for $f_{(N-1)}$. We can integrate these equations into a Lagrangian as
		\begin{eqnarray}
			\cL_{\rm{odd}}^{N} &=& f_{(N-1)} \cdot T(\o) + \frac12 \sum_{n=1}^{N-1} h_{(N-n)} \cdot D(\o) h_{(n)} +  \frac12 \sum_{n=1}^{N-1}  \sum_{m=1}^{n-1} h_{(N-n)} \cdot h_{(m)} \times f_{(n-m)} \nn\\
			&& + \sum_{n=1}^{N-1} e \cdot h_{(N-n)} \times f_{(n)} + f_{(N-1)} \cdot R(\o)  + \frac12 \sum_{n=2}^{N-2} f_{(N-n)} \cdot D(\o) f_{(n-1)} \nn\\ 
			&&	+ \frac16 \sum_{n=2}^{N-1}  \sum_{m=1}^{n-2} f_{(N-n)} \cdot f_{(m)} \times f_{(n-m-1)} \,. \quad 
		\end{eqnarray}
		As in the case of even models, all $N^{th}$-order Lagrangians, including the exotic 
		Einstein-Hilbert term and conformal gravity, can be added to in a sequence without spoiling the structure of field equations
		\begin{eqnarray}
			\cL &=& \frac1{2\mu} \left(\o \cdot d\o + \frac13 \o \cdot \o \times \o \right) + \frac{1}{2} \mu e \cdot T(\o)   + h_{(1)} \cdot T(\o) + \sum_{N} b_N \cL_{\rm{odd}}^N \,,
		\end{eqnarray}
		where $\mu$ is a constant with mass dimension $1$ and $b_N$ are constants with mass dimension $-2N-1$.
		
		\section{Chern--Simons-Like formulation of exotic massive 3D gravity models} \label{S3}
		\paragraph{}
		The Chern--Simons-like models that we provided so far have a corresponding Lagrangian with a single metric. This can be understood by the fact that the exterior derivative of each equation in \eqref{Eom}, or in \eqref{Odd1} and \eqref{Odd2}, vanishes, without referring to the field equation under consideration, itself \cite{Afshar:2019npk}.  In the case of the third-way to 3D gravity, the field equation takes the generic form given in \eqref{3rdWayEOM} where the divergence of the field equation is proportional to the field equation itself \eqref{Div3rdWay}. Consequently, the distinctive feature of their Chern--Simons-like formulation is that the dynamical equation of motion must refer to itself in order for its covariant exterior derivative to vanish \cite{Afshar:2019npk}. Before we proceed with the construction of the Chern--Simons-like Lagrangians, a brief discussion on the properties of the third-way metric equation \eqref{3rdWayEOM} is in order. First of all, the Einstein-like tensor $\mathcal{G}_{\mu\nu}$ does not play any role in the vanishing of the covariant derivative of the field equation \eqref{Div3rdWay}, and can be kept nonetheless.  Furthermore,  we haven't paid attention to the dimensional parameters that should multiply $\mathcal{G}, \mathcal{H}$ and $\mathcal{L}$. However, the divergence equation \eqref{Div3rdWay} indicates that if the parameter that multiplies $\mathcal{H}$ is $\alpha$, then the parameter that multiplies $\mathcal{L}$ must be $\alpha^2$. There is, however, an exception to this rule if $\mathcal{H}$ is the Cotton tensor, in which case the coefficients of $\mathcal{H}$ and $\mathcal{L}$ are independent. This case corresponds to MMG \cite{Bergshoeff:2014pca}. Setting the dimensional constants to unity, third-way equation can be schematically written as 
		\begin{equation}
			0 = R(\omega) + \frac12 e \times e + e \times \mathcal{A} + D(\omega)  \mathcal{A} + \frac12  \mathcal{A} \times  \mathcal{A} \,,
			\label{CSLike3rdWay}
		\end{equation}
		where $\mathcal{A}$ is Lorentz-valued one-form corresponding to a Schouten-like tensor. Based on this equation, we may separate the construction of the Chern--Simons-like formulation of the parity-odd, exotic massive 3D gravity models into two distinctive classes.
		First, we ignore the $e \times  \mathcal{A}$ term, which is the Einstein-like tensor in the metric formulation. The equation of motion \eqref{CSLike3rdWay} indicates that if $\mathcal{A}$ is parity-odd, then the equation of motion is parity-odd as well, see \eqref{Parity} for the parity assignments of the fields. This would enable us to construct a Lagrangian three-form with an odd parity, and the simplest representative of such models is the Exotic Massive Gravity \cite{Ozkan:2018cxj}. In this case, the Schouten-like $\mathcal{A}$ must be formed out of $f_{(n)}$'s only, giving rise to the exotic massive 3D gravity models. The Einstein-like tensor can be recovered by a parity violating modification in the definition of $\mathcal{A}$ 
		\begin{eqnarray}
			\mathcal{A}^a = \sum_n a_n f_{(n)}^a + e^a \,,
			\label{GenOdd}
		\end{eqnarray}
		in which case $\mathcal{A} \times \mathcal{A}$ generates the desired Einstein-like tensors via the $e \times f_{(n)}$ term. The simplest representative of such models is the Exotic General Massive Gravity  \cite{Ozkan:2018cxj}.  In the following subsections, we shall consider each of these cases separately, construct their corresponding Chern--Simons-like formulations and provide rigorous examples.
		\subsection{Chern--Simons-Like formulation}
		The construction of exotic models corresponds to choosing 
		\begin{eqnarray}
			\mathcal{A}^a = \sum_n a_n f_{(n)}^a  \,,
			\label{AParityOdd}
		\end{eqnarray}
		in \eqref{CSLike3rdWay}. Based on the schematic form of the field equation for the third-way to 3D gravity, it is tempting to consider $\mathcal{A}$ as a torsion since defining $\Omega = \omega + \mathcal{A}$ allows us to rewrite \eqref{CSLike3rdWay} as the cosmological Einstein-Hilbert action with torsional connection \cite{Deger:2022gim}
		\begin{eqnarray}
			0 = R(\Omega) + \frac12 e \times e \,.
			\label{Torsional3rdWay}
		\end{eqnarray}
		As we shall elaborate the details momentarily, this is indeed the correct approach if $a_1 = 0$ in \eqref{AParityOdd}. On the other hand, for $a_1 \neq 0$, a better approach is to add the necessary $f_{(1)}$ terms by hand and to consider the following torsional equation
		\begin{eqnarray}
			0 &=& R(\Omega) + \frac12 e \times e + a_1 D(\Omega) f_{(1)} + \frac12 a_1^2 f_{(1)} \times f_{(1)}\,,
			\label{EMGToTorsional3rdWay}
		\end{eqnarray}
		which is basically the EMG with a torsional connection. Here, we set all dimensional constants to unity, and $f_{(1)}$ corresponds to the Cotton tensor in accordance with \eqref{Eom}. Therefore, in both $a_1 =0$ and $a_1 \neq 0$ approaches, we shall consider 
		\begin{eqnarray}
			\mathcal{A}^a = \sum_{n=2}^N a_n f_{(n)}^a  \,,
			\label{ATorsion1}
		\end{eqnarray}
		as a torsion. We now investigate each case separately and discuss their corresponding Chern--Simons-like formulation.

		\subsubsection{Models with \texorpdfstring{$a_1 = 0$}{}}
		For $a_1 =0$, the construction starts with establishing an action principle for a third-way gravity with $f_{(N)}$ chosen to be the torsion. This action is given by
		\begin{align}
			\cL^N  = & \   h_{(N)}  \cdot T(\O) + e \cdot h_{(N)} \times f_{(N)}  + \frac12 \sum_{n=1}^{N-1} \left(h_{(N-n)} \cdot D(\O)h_{(n)}+ h_{(N-n)} \cdot h_{(n)} \times f_{(N)}\right) \nn \\
			& +\sum_{n=1}^{N-1}  e \cdot h_{(N-n)} \times f_{(n)} + \frac12 \sum_{n=1}^{N-1} \sum_{m=1}^{n-1} h_{(N-n)} \cdot h_{(m)} \times f_{(n-m)} +  \frac12 a_N \, e \cdot e\times f_{(N)} \nn  \\
			& + f_{(N-1)} \cdot \left( R(\O) +D(\Omega)f_{(N)} + \frac12 f_{(N)}\times f_{(N)}\right) + \frac12 \gamma  \left(\O \cdot d\O + \frac13 \O \cdot \O \times \O \right)    \nn\\
			&  + \frac12 \sum_{n=1}^{N-2} f_{(N-n-1)} \cdot \left(D(\o) f_{(n)} + f_{(n)} \times f_{(N)}\right)  + \frac16 \sum_{n=1}^{N-2} \sum_{m=1}^{n-1} f_{(N-n-1)}  \cdot f_{(m)} \times f_{(n-m)} \,.
			\label{ExoticAction1}
		\end{align}
		Here, it is convenient to choose the coefficient of the Lorentz-Chern-Simons term as $\gamma = - \ell^2 \alpha_N$, where $\ell$ is the AdS radius, so that AdS is a maximally symmetric solution of the field equations of \eqref{ExoticAction1}. However, this coefficient can be kept arbitrary in general. In order to see that this Lagrangian reproduces the correct solvable set of field equations \eqref{Hierachy-Lambda} as well as the third-way field equation \eqref{CSLike3rdWay}, we consider the field equations of the Lagrangian \eqref{ExoticAction1}
		\begin{eqnarray}
			\delta h_{(N)} &:& 0 = T(\O) + e \times  f_{(N)} \,, \nonumber\\
			\delta e &:& 0= D(\O) h_{(N)} + h_{(N)} \times f_{(N)} + a_N e \times f_{(N)} + \sum_{m=1}^{N-1} h_{(m)} \times f_{(N-m)} \,, \nonumber\\
			\delta h_{(N-n)} &:&	0 = D(\O) h_{(n)} + h_{(n)} \times f_{(N)} + e \times f_{(n)} + \sum_{m=1}^{n-1} h_{(m)} \times f_{(n-m)}  \,,\nonumber\\
			\delta f_{(N-1)} &:&	0 = R(\O) + D(\O) f_{(N)} +\frac12 f_{(N)} \times f_{(N)}+ e \times h_{(1)}\,,\nonumber\\
			\delta f_{(N-n-1)} &:&	0 = D(\O) f_{(n)} +  f_{(n)}\times f_{(N)}+ e \times h_{(n+1)} + \frac12 \sum_{m=1}^{n-1} f_{(m)} \times f_{(n-m)} \nonumber\\
			&& \;\; \quad + \frac{1}2  \sum_{m=1}^{n} h_{(m)} \times h_{(n-m+1)} \,,\nonumber\\ 
			\delta f_{(N)} &:&	0 = e \times h_{(N)} +  \frac12 \sum_{n=1}^{N-1}h_{(N-n)} \times  h_{(n)} + \frac12 a_N e \times e  + D(\O) f_{(N-1)}  \nn \\
			&& \;\; \quad + f_{(N)} \times  f_{(N-1)} + \frac{1}{2} \sum_{n=1}^{N-2} f_{(N-n-1)} \times  f_{(n)} \,,\nonumber\\
			\delta \O &:&	0 = e \times h_{(N)} +  \frac12 \sum_{n=1}^{N-1}h_{(N-n)} \times  h_{(n)}   + D(\O) f_{(N-1)}  +  f_{(N)} \times  f_{(N-1)}    \nonumber\\
			&& \;\; \quad - \ell^2 a_N  R(\O) + \frac{1}{2} \sum_{n=1}^{N-2} f_{(N-n-1)} \times  f_{(n)}  \,,
			\label{TorsionalEOM-Exotic}
		\end{eqnarray}
		where $\delta h_{(N-n)}$ equation holds for $(1 \leq n \leq N-1)$ while $\delta f_{(N-n-1)}$ equation holds for $(1 \leq n \leq N-2)$. First, we note that the last two equations give rise to
		\begin{eqnarray}
			0 = R(\Omega) + \frac{1}{2\ell^2} e \times e \,,
		\end{eqnarray}
		which is the torsional third-way to 3D gravity field equation \eqref{Torsional3rdWay}. Next, upon defining a torsion-free connection via $\Omega = \omega - f_{N}$, we can rewrite the field equations \eqref{TorsionalEOM-Exotic} as
		\begin{eqnarray}
			\delta h_{(N)} &:& 0 = T(\o)  \,, \nonumber\\
			\delta e &:& 0= D(\o) h_{(N)}  + a_N e \times f_{(N)} + \sum_{m=1}^{N-1} h_{(m)} \times f_{(N-m)} \,, \nonumber\\
			\delta h_{(N-n)} &:&	0 = D(\o) h_{(n)}  + e \times f_{(n)} + \sum_{m=1}^{n-1} h_{(m)} \times f_{(n-m)}  \,,\nonumber\\
			\delta f_{(N-1)} &:&	0 = R(\o) + e \times h_{(1)}\,,\nonumber\\
			\delta f_{(N-n-1)} &:&	0 = D(\o) f_{(n)}+ e \times h_{(n+1)} + \frac12 \sum_{m=1}^{n-1} f_{(m)} \times f_{(n-m)} \nonumber\\
			&& \;\; \quad + \frac{1}2  \sum_{m=1}^{n} h_{(m)} \times h_{(n-m+1)} \,,\nonumber\\ 
			\delta f_{(N)} &:&	0 = e \times h_{(N)} +  \frac12 \sum_{n=1}^{N-1}h_{(N-n)} \times  h_{(n)} + \frac12 a_N e \times e  + D(\o) f_{(N-1)}  \nn \\
			&& \;\; \quad  + \frac{1}{2} \sum_{n=1}^{N-2} f_{(N-n-1)} \times  f_{(n)} \,,\nonumber\\
			\delta \o &:&	0 = e \times h_{(N)} +  \frac12 \sum_{n=1}^{N-1}h_{(N-n)} \times  h_{(n)}   + D(\o) f_{(N-1)}  - \ell^2 a_N  R(\o)     \nonumber\\
			&& \;\; \quad + \ell^2 a_N D(\o) f_{(N)} - \frac12 \ell^2 a_N f_{(N)} \times f_{(N)}  + \frac{1}{2} \sum_{n=1}^{N-2} f_{(N-n-1)} \times  f_{(n)}   \,.
		\end{eqnarray}
		These equations are indeed consistent with the solvable set of equations \eqref{Hierachy-Lambda}, except that there is an additional commutator $[P_{a}, P_{b}] = a_N \e_{abc} F^{c(N)}$ that arises for the integrability of the field equations for the given variations. Furthermore, the last two equations give rise to
		\begin{eqnarray}
			0 &=& R(\omega) + \frac1{2\ell^2} e \times e -  D(\omega) f_{(N)} + \frac12 f_{(N)} \times f_{(N)} \,,
		\end{eqnarray}
		which is equation of motion for the third-way to 3D gravity \eqref{CSLike3rdWay} with $\mathcal{A} = - f_{(N)}$. As evident from the $\delta e$ equation, the model can be extended to include all $a_n$ with $n \neq 1$ in the definition of $\mathcal{A}$ with the following simple modification
		\begin{eqnarray}
			\cL_{a_1=0} = \cL^N + \frac12 \sum_{n=2}^{N-1} a_n e \cdot e \times f_{(n)} \,,
			\label{ExoticAction2}
		\end{eqnarray}
		where the AdS condition still fixes the coefficient of the Lorentz-Chern-Simons term $\gamma$ in \eqref{ExoticAction1} to be $\gamma = -\ell^2 a_N$. With this modification, the $\delta e$ equation is given by
		\begin{eqnarray}
			\delta e &:& 0= D(\o) h_{(N)}  + a_N e \times f_{(N)} + \sum_{n=2}^{N-1} a_n e \times f_{(n)} + \sum_{m=1}^{N-1} h_{(m)} \times f_{(N-m)} \,,
			\label{DeltaE}
		\end{eqnarray}
		while the $\delta f_{(n)}$ ($n \neq 1$) equations are modified by a simple $e \times e$ term that does not affect the solvability of the field equations. The modified $\delta e$ equation \eqref{DeltaE} implies that $f_{(N)}$ now contains its previous definition as well as all $f_{(n)}$'s with $1 < n < N$, consequently giving rise to \eqref{ATorsion1}. 
		
		At this stage, it is worthwhile to discuss why $a_1$ is not included in \eqref{ExoticAction1}. If $a_1 f_{(1)}$ is included in the sum, i.e.
		\begin{eqnarray}
			\cL = \cL^N + \frac12 \sum_{n=2}^{N-1} a_n e \cdot e \times f_{(n)} + \frac12 a_1 e \cdot e \times f_{(1)}\,,
			\label{ExoticAction3}
		\end{eqnarray}
		it can be removed by considering the following simultaneous redefinition of the fields in \eqref{ExoticAction3}
		\begin{align}
			N \neq 2 &:& h_{(N-1)} & \to   h_{(N-1)} - \frac12 a_1 e\,, & h_{(N)} & \to h_{(N)} + \frac12 a_1 h_{(1)} \,,\nonumber\\
			N = 2 &:& h_{(1)} & \to h_{(1)} - \frac12 a_1 e \,,  &  h_{(2)} & \to h_{(2)} + \frac12 a_1 h_{(1)} - \frac18 a_1^2 e \,.
		\end{align}
		Hence, as mentioned, including the $f_{(1)}$ term in the third-way action cannot be achieved with this construction, and should be considered separately. 
		
		We may, however, include the Einstein-like tensor in the third-way field equation, which corresponds to choosing $\mathcal{A}$ as in \eqref{GenOdd}, leading to exotic general models of massive 3D gravity. This can be achieved by including a cosmological constant in the action principle
		\begin{eqnarray}
			\cL = \cL^N + \frac12 \sum_{n=2}^{N-1} a_n e \cdot e \times f_{(n)} + \frac16 \Lambda_0 \, e \cdot e \times e \,.
			\label{ExoticAction4}
		\end{eqnarray}
		In this case, the torsion-free equations are given by
		\begin{eqnarray}
			\delta h_{(N)} &:& 0 = T(\o)  \,, \nonumber\\
			\delta e &:& 0= D(\o) h_{(N)}  + a_N e \times f_{(N)} + \sum_{n=2}^{N-1} a_n e \times f_{(n)} + \sum_{m=1}^{N-1} h_{(m)} \times f_{(N-m)} \nonumber\\
			&& \;\; \quad + \frac12 \L_0 e \times e \,, \nonumber\\
			\delta h_{(N-n)} &:&	0 = D(\o) h_{(n)}  + e \times f_{(n)} + \sum_{m=1}^{n-1} h_{(m)} \times f_{(n-m)}  \,,\nonumber\\
			\delta f_{(N-1)} &:&	0 = R(\o) + \frac12 a_{N-1} e \times e + e \times h_{(1)}\,,\nonumber\\
			\delta f_{(N-n-1)} &:&	0 = D(\o) f_{(n)} + \frac12 a_{N-n-1} e \times e + e \times h_{(n+1)} + \frac12 \sum_{m=1}^{n-1} f_{(m)} \times f_{(n-m)} \nonumber\\
			&& \;\; \quad + \frac{1}2  \sum_{m=1}^{n} h_{(m)} \times h_{(n-m+1)} \,,\nonumber\\ 
			\delta f_{(N)} &:&	0 = e \times h_{(N)} +  \frac12 \sum_{n=1}^{N-1}h_{(N-n)} \times  h_{(n)} + \frac12 a_N e \times e  + D(\o) f_{(N-1)}  \nn \\
			&& \;\; \quad  + \frac{1}{2} \sum_{n=1}^{N-2} f_{(N-n-1)} \times  f_{(n)} \,,\nonumber\\
			\delta \o &:&	0 = e \times h_{(N)} +  \frac12 \sum_{n=1}^{N-1}h_{(N-n)} \times  h_{(n)}   + D(\o) f_{(N-1)}  + \gamma  R(\o)     \nonumber\\
			&& \;\; \quad - \gamma D(\o) f_{(N)} + \frac12 \gamma f_{(N)} \times f_{(N)}  + \frac{1}{2} \sum_{n=1}^{N-2} f_{(N-n-1)} \times  f_{(n)}   \,,
		\end{eqnarray}
		where $a_1 = 0$ and $\delta h_{(N-n)}$ equation holds for $(1 \leq n \leq N-1)$ while $\delta f_{(N-n-1)}$ equation holds for $(1 \leq n \leq N-2)$. The coefficient of the Lorentz-Chern-Simons term, $\gamma$, can be fixed by setting the background values
		\begin{align}
			\bar{h}_{(n)} & = c_n \bar{e} \,, & \bar{f}_{(n)} & = b_n \bar{e} \,,
		\end{align}
		and demanding that AdS is a solution to this set of field equations. This condition allows us to determine $c_1$ via the $\delta f_{(N-1)}$ field equation as 
		\begin{eqnarray}
			c_1 = \frac12 \left( \frac{1}{\ell^2} - a_{N-1} \right) \,,
		\end{eqnarray}
		Furthermore, the $\delta h_{(N-n)}$ equation fixes $b_n =0$ for $n \neq N$, and $\delta f_{(N-n-1)}$ and $\delta f_{(N-1)}$ determines the values for $c_{n}$ in terms of the free parameters of the theory and the AdS radius $\ell$ as
		\begin{align}
			c_{n+1} & = - \frac12 a_{N-n-1} - \frac12 \sum_{m=1}^{n} c_{m} c_{n-m+1} & (1 \leq n \leq N-2) \,,\nonumber\\
			c_N &= - \frac12 \sum_{n=1}^{N-1} c_{N-n} c_{n} - \frac12 a_N \,.
		\end{align}
		The $\delta h_{(N-n)}$ equation implies that $b_n$ for $n < N$ are zero, while $\delta e$ equation can be used to express $b_N$ in terms of $\Lambda_0$ and $a_N$ as
		\begin{equation}
			b_N = - \frac{\Lambda_0}{2a_N} \,. 
		\end{equation}
		Finally, using these results, the $\delta \o$ equation can be used to determine the coefficient of the Lorentz-Chern-Simons term in terms of the other parameters of the theory as
		\begin{equation}
			\gamma  = \frac{4 \ell^2 a_N^3}{\ell^2 \Lambda_0^2 - 4 a_N^2 } \,,
			\label{GenGamma}
		\end{equation}
		which, for $\Lambda_0 =0$, recovers the previous parameterization $\gamma = - a_N \ell^2$. 

  Before ending this section, let us briefly discuss the consequences of fixing $\gamma = - a_N \ell^2$ for the exotic models, where asymptotic Virasoro $\oplus$ Virasoro symmetry algebra implied by the Brown-Henneaux boundary conditions have the central charges $c_\pm \propto \pm \gamma$ \cite{Bergshoeff:2019rdb}. If $\gamma = - a_N \ell^2$, then we may only set $c_\pm = 0$ by setting $a_N = 0$, which is a problem as we no longer have a solvable model in that case. As we shall demonstrate for $a_1 \neq 0$ models, the torsion-free construction leaves the coefficient of the Lorentz-Chern-Simons term free, allowing $c_\pm =0$.

		\subsubsection{Models with \texorpdfstring{$a_1 \neq 0$}{}} \label{SS:A1Neq0}
		As previously mentioned, the case where $a_1 \neq 0$ necessitates using the torsional EMG equation \eqref{EMGToTorsional3rdWay} instead of the torsional cosmological Einstein-Hilbert action \eqref{Torsional3rdWay}. Moreover, torsional models determine the coefficient of the Lorentz-Chern-Simons term, whereas in a torsion-free construction, this coefficient remains an independent parameter. Therefore, we aim to examine the $a_1 \neq 0$ case from a torsion-free perspective to highlight this difference. Additionally, we will demonstrate that transitioning to a torsional framework leads to the torsional EMG equation as the resulting dynamical field equation. This torsion-free exotic Lagrangian is given by
		\begin{align}
		\cL_{a_1 \neq 0}^N = & \, h_{(N)}  \cdot T(\o)  + \frac12 \sum_{n=1}^{N-1} h_{(N-n)} \cdot D(\o)h_{(n)}   +\sum_{n=1}^{N-1}  a_{n} e \cdot h_{(N-n)} \times f_{(n)}   \nn \\
		&+ \frac12 \sum_{n=1}^{N-1} \sum_{m=1}^{n-1} h_{(N-n)} \cdot h_{(m)} \times f_{(n-m)}  + f_{(N-1)} \cdot  R(\o)  + \frac1{2\ell^2} \sum_{n=2}^N e \cdot e \times f_{(n)}  \nn  \\
		&  + \frac12 \sum_{n=1}^{N-2} f_{(N-n-1)} \cdot D(\o) f_{(n)}  + \frac16 \sum_{n=1}^{N-2} \sum_{m=1}^{n-1} f_{(N-n-1)}  \cdot f_{(m)} \times f_{(n-m)} \nn\\
				&  + \frac12 \gamma  \left(\o \cdot d\o + \frac13 \o \cdot \o \times \o \right) + f_{(N)} \cdot R(\o) + \frac12 f_{(N)} \cdot D(\o) f_{(N)}+ \frac16 f_{(N)}  \cdot f_{(N)} \times f_{(N)}  \nn\\
				&  + f_{(N)} \cdot  D(\o) f_{(1)} + \frac12 f_{(1)} \cdot f_{(1)} \times f_{(N)} +\frac12 f_{(1)} \cdot f_{(N)} \times f_{(N)} + f_{(1)} \cdot R(\o)  \nn\\
				&+ \frac12 f_{(1)} \cdot D(\o) f_{(1)} + \frac16  f_{(1)} \cdot f_{(1)} \times f_{(1)} \,,
    \label{TorsionFreeConstruction}
		\end{align}
 where the field equations read
  \begin{eqnarray}
      \delta h_{(N)} &:& 0 = T(\o)\,,  \nonumber\\
        \delta h_{(N-n)} &:&	0 = D(\o) h_{(n)}  +a_n e \times f_{(n)} + \sum_{m=1}^{n-1} h_{(m)} \times f_{(n-m)}  \,,\nonumber\\
        \delta e &:& 0= D(\o) h_{(N)}  + \frac{1}{\ell^2} \sum_{n=2}^N e \times f_{(n)} + \sum_{n=1}^{N-1} a_n h_{(N-n)} \times f_{(n)} \,, \nonumber\\
        \delta f_{(N)} &:& 0=R(\o)+ \frac{1}{2\ell^2} e \times e + D(\o)f_{(N)} + \frac{1}{2} f_{(N)} \times f_{(N)} +D(\o)f_{(1)} \nn\\
        &&  \quad \quad  + \frac{1}{2} f_{(1)} \times f_{(1)} + f_{(1)} \times f_{(N)} \,, \nn \\
        \delta f_{(N-1)} &:& 0= R(\o) + a_{N-1} e \times h_{(1)} + \frac{1}{2\ell^2} e \times e \,, \nn \\
        \delta f_{(N-n-1)} &:&	0 = D(\o) f_{(n)} + \frac{1}{2\ell^2} e \times e  +a_{N-n-1} e \times h_{(n+1)} + \frac12 \sum_{m=1}^{n-1} f_{(m)} \times f_{(n-m)} \nn\\
        && \quad\quad+ \frac{1}2  \sum_{m=1}^{n} h_{(m)} \times h_{(n-m+1)} \,,\nonumber \\
        \delta f_{(1)} &:& 0= D(\o)f_{(N-2)} + a_1 e \times h_{(N-1)} + \frac{1}{2} \sum_{n=1}^{N-2} h_{(n)} \times h_{(N-n-1)} \nonumber\\
        &&  \quad \quad  + \frac{1}{2} \sum_{n=1}^{N-3} f_{(n)} \times f_{(N-n-2)} + R(\o) + D(\o)f_{(N)} + \frac{1}{2} f_{(N)} \times f_{(N)}  \nn \\
        && \quad \quad + D(\o)f_{(1)} + \frac{1}{2}f_{(1)} \times f_{(1)} + f_{(N)} \times f_{(1)} \,, \nn \\
        \delta \o &:&	0 = e \times h_{(N)} +  \frac12 \sum_{n=1}^{N-1}h_{(N-n)} \times  h_{(n)} + \frac{1}{2} \sum_{n=1}^{N-2} f_{(N-n-1)} \times  f_{(n)}   \nonumber\\
        && \quad \quad + D(\o) f_{(N-1)}   +  \gamma  R(\o)  +D(\o) f_{(N)} + \frac12  f_{(N)} \times f_{(N)} + D(\o)f_{(1)}   \nonumber\\
			&&  \quad \quad  + \frac{1}{2} f_{(1)} \times f_{(1)} +f_{(1)} \times f_{(N)}   \,.
   \label{a1non0Exotic}
  \end{eqnarray}
A quick glance shows that $\delta f_{(1)}$ equation contains almost all of the $\delta f_{(N)}$ equation. 
Similarly, the same can be said for the $\delta \omega$ equation up to a factor of $\gamma$ in front of $R(\o)$, which does not spoil the solvability of the field equations. These provide a minor deviation in the field equations from the infinite-dimensional algebra perspective \eqref{Hierachy-Lambda}. As already advertised, $\gamma$ remains an independent parameter. Furthermore, the $\delta f_{(N)}$ equation, which is the dynamical field equation, is the torsional EMG equation \eqref{EMGToTorsional3rdWay} upon setting $\omega = \Omega - f_{(N)}$. Moreover, note that $f_{(N)}$ includes all $f_{(n)}$ with $n < N$ as can be seen from the $\delta e$ equation. Thus, \eqref{TorsionFreeConstruction} can be considered as the $a_1 \neq 0$ equivalent of \eqref{ExoticAction2}. If this is not desired, then the $\sum_{n=2}^N e \cdot e \times f_{(n)}$ piece can simply be replaced by $e \cdot e \times f_{(N)}$. Note also that the lower bound for this sum is $n=2$ rather than $n=1$ since $e \cdot e \times f_{(1)}$ can be removed by a field redefinition as before. Finally, the exotic generalized extension can be obtained by adding a cosmological constant term, i.e. $\Lambda_0 e \cdot e \times e$ as in \eqref{ExoticAction4}. In this case, however, $\ell$ no longer represents the AdS radius, and is modified by a $\L_0$ term as we shall demonstrate explicitly for the $N=2$ case in the next subsection.

		\subsection{\texorpdfstring{$N=2$}{} Exotic general massive gravity models}
		We have now paved the way to discuss the $N \geq 2$  exotic and exotic general models. Here, as an example, we will provide a detailed discussion on the simplest $N=2$ case, including its $a_1 =0$ and $a_1 \neq 0$ formulations and its general exotic extensions. 
		
		For $a_1 = 0$, the general exotic model is given by
		\begin{eqnarray}
			\cL^{(2)}_{a_1=0} &=& f_{(1)} \cdot R(\O)  + h_{(2)} \cdot T(\O)  + f_{(1)} \cdot D(\O) f_{(2)} + \frac12 h_{(1)} \cdot D(\O) h_{(1)} + \frac{a_2}{2} e \cdot e \times f_{(2)} \nn\\
			&& + \frac12 f_{(1)} \cdot f_{(2)} \times f_{(2)}  + e \cdot f_{(1)} \times h_{(1)} + \frac12 f_{(2)} \cdot h_{(1)} \times h_{(1)} + e \cdot f_{(2)} \times h_{(2)} \nonumber\\
			&& + \frac{1}2 \g \left(\O \cdot d\O + \frac13 \O \cdot \O \times \O \right) + \frac16 \L_0 e \cdot e\times e \,,
			\label{N1A10}
		\end{eqnarray}
		where
		\begin{equation}
			\gamma  = \frac{4 \ell^2 a_2^3}{\ell^2 \Lambda_0^2 - 4 a_2^2 } \,,
			\label{N2Gamma1}
		\end{equation}
		in accordance with \eqref{GenGamma}. Here, $\gamma$ is dimensionless while $a_2$ has mass dimension $2$ and $\Lambda_0$ has the mass dimension $3$. Consequently, the overall Lagrangian needs to be multiplied with a constant of mass dimension $-1$, which we ignore here as this does not alter the field equations. The field equations for $h_{(2)}, h_{(1)}$ and $e$ read
		\begin{eqnarray}
			0 &=& T(\O) + e \times f_{(2)}\,,\nn\\
			0 &=& D(\O) h_{(1)} + f_{(2)} \times h_{(1)} + e \times f_{(1)} \,,\nn\\
			0 &=& D(\O) h_{(2)} + f_{(2)} \times h_{(2)}+  f_{(1)} \times h_{(1)} + a_2 e \times f_{(2)} + \frac12 \L_0  e \times e \,.
   \label{N2a10Set1}
		\end{eqnarray}
		These equations are the standard equations that solve $f_{(1)}$ and $f_{(2)}$ in terms of $h_{(1)}, h_{(2)}$ and $e$ in the presence of torsion, $f_{(2)}$. In particular, the first one implies that the spin connection is torsional, i.e. one can set $\Omega = \omega - f_{(2)}$ where $\o$ is the torsion-free spin connection. Next, we turn to the field equations for $f_{(1)}$ and $f_{(2)}$
		\begin{eqnarray}
			0 &=& R(\O) + D(\O) f_{(2)} + \frac12 f_{(2)} \times f_{(2)} + e \times h_{(1)} \,,\nn\\ 
			0 &=& D(\O) f_{(1)} + f_{(1)} \times f_{(2)} + \frac12 h_{(1)} \times h_{(1)} + \frac12  a_2 e \times e +  e \times h_{(2)} \,.
		\end{eqnarray}
		These equations together give rise to the following expressions for $h_{(1)}, h_{(2)}, f_{(1)}$ and $f_{(2)}$  \cite{Afshar:2014ffa}
		\begin{eqnarray}
			h_{(1)\mu\nu} &=&  - S_{\mu\nu} \,, \nn\\
			f_{(1)\mu\nu} & =&  C_{\mu\nu} \,,\nn\\
			h_{(2)\mu\nu}&=& -  D_{\mu\nu} +   \left( P_{\mu\nu} - \frac14 P g_{\mu\nu} \right) + a_2 g_{\m\n} \,, \nn\\
			f_{(2)\mu\nu} & =&  \frac{1}{a_2} \left(  - E_{\mu\nu} - 2 \left(Q_{\mu\nu} - \frac14 Q g_{\mu\nu} \right) + S C_{\mu\nu} +  \L_0 g_{\mu\nu} \right) \,, 
			\label{TensorDefinition}
		\end{eqnarray}
		where $S_{\mu\nu}$ and $ C_{\mu\nu}$ are the Schouten and the Cotton tensors, respectively,
		\begin{align}
			S_{\mu\nu} & = R_{\mu\nu} - \frac14 g_{\mu\nu} R \,, &  C_{\mu\nu} & = \e_\m{}^{\r\s} \nabla_\r S_{\n\s} \,, 
		\end{align}
		while $D_{\mu\nu},E_{\mu\nu},P_{\mu\nu}$ and $Q_{\mu\nu}$ are defined by
		\begin{align}
			D_{\mu\nu} & = \e_{\m}{}^{\r\s} \nabla_\r  C_{\n\s} \,, & P_{\mu\nu} & = G_{\m}{}^\l S_{\n\l} \,,\nn\\
			E_{\m\n} & = \e_{\m}{}^{\r\s} \nabla_\r h_{(2)\n\s} \,, & Q_{\m\n} & =  C_{(\m}{}^\l S_{\n)\l} \,.
		\end{align}
		Finally, the $\O$ field equation is given by
		\begin{equation}
			0 = e \times h_{(2)} +  \frac12 h_{(1)} \times  h_{(1)}   + D(\O) f_{(1)}  +  f_{(2)} \times  f_{(1)}  + \g  R(\O) \,,
		\end{equation}
		which, when combined with the $f_{(2)}$ equation, gives rise to
		\begin{eqnarray}
			0 &=& R(\O) - \frac{a_2}{2\g} e \times e \,,
		\end{eqnarray}
		where $\gamma$ is as defined in \eqref{N2Gamma1}. Upon extracting the torsion part of the spin-connection, this would give rise to the field equation for the $N=2$ exotic general massive gravity
		\begin{eqnarray}
			0 &=& G_{\mu\nu} +  \frac{a_2}{\g} g_{\mu\nu} - \epsilon_{\m}{}^{\r\s} \nabla_\r {f}_{(2)\n\s} + \frac12 \e_{\m}{}^{\r\s} \e_\n{}^{\l\t} {f}_{(2)\r\l} {f}_{(2)\s\t} \,.
   \label{a10EOM}
		\end{eqnarray}
		As expected, this model has two free parameters $(a_2, \L_0)$. 

  Next, we focus on the $a_1 \neq 0$ case, where the $N=2$ action is given by
  \begin{align}
      \cL_{a_1 \neq 0}^{(2)} =& \, h_{(2)} \cdot T(\o) + \frac12 h_{(1)} \cdot D(\omega) h_{(1)} + a_1 e \cdot f_{(1)} \times h_{(1)} + \frac{1}{2\ell^2} e \cdot e \times f_{(2)}  \nn\\
      & +2 f_{(1)} \cdot R(\omega)  + \frac12 \gamma \left( \omega \cdot d\omega + \frac13 \omega \cdot \omega \times \omega \right) + \frac12 f_{(1)} \cdot D(\omega) f_{(1)} + f_{(1)} \cdot D(\omega) f_{(2)} \nn\\
      &  + \frac12 f_{(2)} \cdot D(\omega) f_{(2)} + f_{(2)} \cdot R(\omega) + \frac16 f_{(1)} \cdot f_{(1)} \times f_{(1)} + \frac12 f_{(1)} \cdot f_{(1)} \times f_{(2)} \nn\\
      & + \frac12 f_{(1)} \cdot f_{(2)} \times f_{(2)}  + \frac16 f_{(2)} \cdot f_{(2)} \times f_{(2)} + \frac16 \Lambda_0 e \cdot e \times e \,,
  \end{align}
	in accordance with \eqref{a1non0Exotic}. Once again, $\gamma$ is dimensionless, $\Lambda_0$ has the mass dimension $3$ and $a_1$ has mass dimension $1$. Consequently, the overall Lagrangian needs to be multiplied with a constant of mass dimension 1, which we ignore as this does not alter the field equations. This model is identical to the $N=2$ exotic theory introduced in \cite{Afshar:2019npk} except for the Lorentz-Chern-Simons term, implying that the construction of \cite{Afshar:2019npk} corresponds to the critical $(\g = 0)$ $N=2$ exotic model with $a_1 \neq 0$, where the central charges vanish. The field equations for $h_{(2)}, h_{(1)}$ and $e$ read
 \begin{eqnarray}
    0 &=& T(\omega) \,,\nn\\
    0 &=& D(\o) h_{(1)} + a_1 e \times f_{(1)} \,,\nn\\
    0 &=& D(\omega) h_{(2)} + a_1 f_{(1)} \times h_{(1)}  + \frac{1}{\ell^2} e \times f_{(2)} + \frac12 \L_0 e \times e \,,
 \end{eqnarray}
     which are the analogs of \eqref{N2a10Set1}. These equations allow us to identify $\omega$ to be the torsion-free spin connection and solve $f_{(1)}$ and $f_{(2)}$ in terms of $h_{(2)}, h_{(1)}$ and $e$. Next, we consider $f_{(2)}, f_{(1)}$ and $\omega$ field 
     equations
\begin{eqnarray}
0 &=& R(\o) + \frac1{2\ell^2} e \times e + D(\o) f_{(1)} + D(\o) f_{(2)} + \frac12 f_{(1)} \times f_{(1)} + \frac12 f_{(2)} \times f_{(2)} + f_{(1)} \times f_{(2)} \,, \nn\\ 
    0 &=& a_1 e \times h_{(1)} + 2 R(\o) + D(\o) f_{(1)} + D(\o) f_{(2)} + \frac12 f_{(1)} \times f_{(1)}   + \frac12 f_{(2)} \times f_{(2)} + f_{(1)} \times f_{(2)} \,, \nn\\
    0 &=& e \times h_{(2)} + \frac12 h_{(1)} \times h_{(1)} + 2 D(\o) f_{(1)} + \g R(\o)  + D(\o) f_{(2)} + \frac12 f_{(1)} \times f_{(1)} + f_{(1)} \times f_{(2)} \nn\\
    && + \frac12 f_{(2)} \times f_{(2)}  \,. 
\end{eqnarray}
    Here, the first equation is the dynamical field equation which is of the desired form. The second and the third equations can be simplified by using the $\delta f_{(2)}$ field equation as
    \begin{eqnarray}
        0 &=& R(\o) - \frac1{2\ell^2} e \times e + a_1 e \times h_{(1)} \,,\nn\\
        0 &=& D(\o) f_{(1)} + \frac12 h_{(1)} \times h_{(1)} - \frac1{2\ell^2} e \times e  + (\g-1) R(\o) + e \times h_{(2)} \,.
    \end{eqnarray}
It is convenient to make the following simultaneous field redefinition to determine the explicit form of the auxiliary fields
\begin{align}
h_{(1)} & \to H_{(1)} + \frac{e}{2a_1 \ell^2} \,, & h_{(2)} & \to \left( -a_1  + a_1 \g - \frac{1}{2\ell^2 a_1} \right) H_{(1)} + H_{(2)} + \left(\frac{1}{2\ell^2} - \frac{1}{8 \ell^4 a_1^2} \right) e \,,\nn\\
f_{(1)} & \to F_{(1)} \,, & f_{(2)} & \to \ell^2 F_{(2)} + \left(  -1 - \ell^2 a_1^2+ \ell^2 a_1^2 \g \right) F_{(1)} - \frac{\ell^2 \L_0}{2} e \,.
\end{align}
In terms of these new variables, the field equations are given by
\begin{eqnarray}
    0 &=& D(\o) H_{(1)} + a_1 e \times F_{(1)} \,,\nn\\
    0 &=& D(\o) H_{(2)} + a_1 F_{(1)} \times H_{(1)} + e \times F_{(2)}\,,\nn\\
    0 &=& R(\omega) + a_1 e \times H_{(1)} \,,\nn\\
    0 &=& D(\o) F_{(1)} + \frac12 H_{(1)} \times H_{(1)} + e \times H_{(2)} \,,\nn\\ 
    0 &=& R(\o) + \frac12 \left( \frac{1}{\ell^2} + \frac{1}{4}\ell^4 \L_0^2\right) e \times e  + \a D(\o) F_{(1)} + \ell^2 D(\o) F_{(2)} + \frac12 \a^2 F_{(1)} \times F_{(1)} \nn\\
    && + \frac12 \ell^4 F_{(2)} \times F_{(2)} + \a \ell^2 F_{(1)} \times F_{(2)} - \frac12 \a \ell^2 \L_0  e \times F_{(1)} - \frac12 \L_0 \ell^4 e \times F_{(2)} \,,
\end{eqnarray}
where we exclude the torsion constraint equation as it is not affected by the change of variables, and define $\a = -\ell^2 a_1^2 (1 - \g)$. Note that for $\L_0 = 0$, we have the standard dynamical field equation for the exotic $N=2$ theory. These equations imply the following expressions for $H_{(1)}, H_{(2)}, F_{(1)}$ and $F_{(2)}$  \cite{Afshar:2014ffa}
		\begin{align}
			H_{(1)\mu\nu} &=  - \frac1{a_{1}} S_{\mu\nu} \,, & H_{(2)\mu\nu}&=  \frac1{a_{1}^2} \left( D_{\mu\nu} - \left( P_{\mu\nu} - \frac14 P g_{\mu\nu} \right) \right) \,, \nn\\
			F_{(1)\mu\nu} & = \frac1{a_{1}^2} C_{\mu\nu} \,, & F_{(2)\mu\nu} & =  \frac{1}{a_1^2} \left(  - E_{\mu\nu} - 2 \left(Q_{\mu\nu} - \frac14 Q g_{\mu\nu} \right) + S C_{\mu\nu} \right) \,.
			\label{TensorDefinition2}
		\end{align}
Consequently, the dynamical equation of motion can be written as
\begin{eqnarray}
    0 &=& G_{\m\n} - \left( \frac{1}{\ell^2} + \frac{1}{4}\ell^4 \L_0^2\right) g_{\m\n} + \alpha \e_\m{}^{\r\s} \nabla_\r F_{(1) \n\s} + \frac12 \a^2 \e_{\m}{}^{\r\s} \e_\n{}^{\l\t} {F}_{(1)\r\l} {F}_{(1)\s\t}  \nn\\
    && + \ell^2 \e_\m{}^{\r\s} \nabla_\r F_{(2) \n\s}  + \frac12 \ell^4 \e_{\m}{}^{\r\s} \e_\n{}^{\l\t} {F}_{(2)\r\l} {F}_{(2)\s\t} + \a \ell^2  \e_{\m}{}^{\r\s} \e_\n{}^{\l\t} {F}_{(1)\r\l} {F}_{(2)\s\t} \nn\\
    && + \frac12 \a \ell^2 \L_0 \left( {F}_{(1)\m\n} - g_{\m\n} g^{\r\l} {F}_{(1)\r\l}\right) +  \frac12  \ell^4 \L_0 \left( {F}_{(2)\m\n} - g_{\m\n} g^{\r\l} {F}_{(2)\r\l}\right) \,,
    \label{a1Non0Eom}
\end{eqnarray}
which now includes the $f_{(1)}$ terms that were missing in the previous construction \eqref{a10EOM}.
		\section{Conclusions} \label{Conclusions}
\paragraph{}
  In this paper, we have considered the Chern--Simons-like formulation of the exotic and exotic general massive three-dimensional gravity models. The construction has mainly been divided in two subclasses based on the generic formulation of the third-way to 3D gravity field equation \eqref{CSLike3rdWay}. First, when $\mathcal{A}$ does not include $f_{(1)}$,  the construction can be achieved with a torsional cosmological Einstein equation, where the generic Lagrangian three-form is presented in \eqref{ExoticAction3}. This torsional interpretation, though mathematically appealing, fixes the coefficient of the Lorentz-Chern-Simons term, preventing the construction of critical exotic models where the central charges vanish. Conversely, when $\mathcal{A}$ includes $f_{(1)}$, the torsional construction is accomplished by considering the torsional exotic massive 3D gravity, rather than the cosmological Einstein equation. Critical models become accessible by forgoing the torsional interpretation, as detailed in the generic form of the exotic Lagrangian three-form presented in \eqref{TorsionFreeConstruction}. We have also presented the $N=2$ exotic general model as an explicit example, providing both its torsional and torsion-free constructions.

  This paper represents the initial step toward the Chern--Simons-like formulation of third-way to three-dimensional gravity models. The next logical step is to extend our work to parity-mixing models, where the leading-order model is the Minimal Massive Gravity (MMG). In such a case, $\mathcal{A}$ would presumably be given by
\begin{equation}
\mathcal{A} = \sum_{n=1}^N \left( a_n f_{(n)} + b_{n} h_{(n)} \right)  + e \,.
\end{equation} 
The construction of this model would allow for the identification of a class of third-way models that avoid bulk/boundary unitarity issues. However, it is important to note that MMG's ability to achieve unitarity stems from its inclusion of an extra parameter in its field equation. This is because $\mathcal{A}$ is identified as the Schouten tensor, and $D(\o)\mathcal{A}$, corresponding to the Cotton tensor, is divergence-free, leaving the coefficient of $\frac12 \mathcal{A} \times \mathcal{A}$ free in \eqref{CSLike3rdWay}. This freedom is not present in a generic case, resulting in fewer parameters than the number of constraints imposed by bulk/boundary unitarity of the theory.

Our construction is primarily based on the recent observation that an infinite-dimensional algebra underlies the massive 3D gravity models compatible with a holographic c-theorem \cite{Bergshoeff:2021tbz}. This algebra is closely related to the Lie algebra expanded AdS$_3$ algebra. Given the non-relativistic relation between Lie algebra expansion and the contraction of multiple copies of the Poincar\'e (or AdS) algebra \cite{Ekiz:2022wbi}, we expect that the infinite-dimensional algebra underlying 3D gravity models relates to multiple copies of the AdS algebra. Explicitly demonstrating this relationship would connect multi-metric gravity models and 3D massive gravity theories, a connection shown so far only for a few cases \cite{Bergshoeff:2013xma, Afshar:2014dta, Ozkan:2019iga}. 

Another interesting continuation of our work would be to extend the infinite-dimensional algebra to an infinite-dimensional superalgebra. So far, the only third-way supergravity has been achieved for MMG \cite{Deger:2022gim,Deger:2023eah}, which was obtained up to quartic-fermions. With the methodology we provide here, the algebra viewpoint might be helpful to obtain supersymmetry in 
third-way models to all orders. We hope to address this question in the near future.

		\bibliographystyle{utphys.bst}
		\bibliography{ref}

\providecommand{\href}[2]{#2}\begingroup\raggedright\begin{thebibliography}{10}

\bibitem{Banados:1992wn}
M.~Banados, C.~Teitelboim, and J.~Zanelli, ``{The Black hole in
  three-dimensional space-time},''
  \href{http://dx.doi.org/10.1103/PhysRevLett.69.1849}{{\em Phys. Rev. Lett.}
  {\bfseries 69} (1992) 1849--1851},
  \href{http://arxiv.org/abs/hep-th/9204099}{{\ttfamily arXiv:hep-th/9204099}}.

\bibitem{Banados:1992gq}
M.~Banados, M.~Henneaux, C.~Teitelboim, and J.~Zanelli, ``{Geometry of the
  (2+1) black hole},'' \href{http://dx.doi.org/10.1103/PhysRevD.48.1506}{{\em
  Phys. Rev. D} {\bfseries 48} (1993) 1506--1525},
  \href{http://arxiv.org/abs/gr-qc/9302012}{{\ttfamily arXiv:gr-qc/9302012}}.
  [Erratum: Phys.Rev.D 88, 069902 (2013)].

\bibitem{Brown:1986nw}
J.~D. Brown and M.~Henneaux, ``{Central Charges in the Canonical Realization of
  Asymptotic Symmetries: An Example from Three-Dimensional Gravity},''
  \href{http://dx.doi.org/10.1007/BF01211590}{{\em Commun. Math. Phys.}
  {\bfseries 104} (1986) 207--226}.

\bibitem{Achucarro:1986uwr}
A.~Achucarro and P.~K. Townsend, ``{A Chern-Simons Action for Three-Dimensional
  anti-De Sitter Supergravity Theories},''
  \href{http://dx.doi.org/10.1016/0370-2693(86)90140-1}{{\em Phys. Lett. B}
  {\bfseries 180} (1986) 89}.

\bibitem{Witten:1988hc}
E.~Witten, ``{(2+1)-Dimensional Gravity as an Exactly Soluble System},''
  \href{http://dx.doi.org/10.1016/0550-3213(88)90143-5}{{\em Nucl. Phys. B}
  {\bfseries 311} (1988) 46}.

\bibitem{Hohm:2012vh}
O.~Hohm, A.~Routh, P.~K. Townsend, and B.~Zhang, ``{On the Hamiltonian form of
  3D massive gravity},''
  \href{http://dx.doi.org/10.1103/PhysRevD.86.084035}{{\em Phys. Rev. D}
  {\bfseries 86} (2012) 084035},
  \href{http://arxiv.org/abs/1208.0038}{{\ttfamily arXiv:1208.0038 [hep-th]}}.

\bibitem{Bergshoeff:2014bia}
E.~A. Bergshoeff, O.~Hohm, W.~Merbis, A.~J. Routh, and P.~K. Townsend,
  ``{Chern-Simons-like Gravity Theories},''
  \href{http://dx.doi.org/10.1007/978-3-319-10070-8_7}{{\em Lect. Notes Phys.}
  {\bfseries 892} (2015) 181--201},
  \href{http://arxiv.org/abs/1402.1688}{{\ttfamily arXiv:1402.1688 [hep-th]}}.

\bibitem{Afshar:2014ffa}
H.~R. Afshar, E.~A. Bergshoeff, and W.~Merbis, ``{Extended massive gravity in
  three dimensions},'' \href{http://dx.doi.org/10.1007/JHEP08(2014)115}{{\em
  JHEP} {\bfseries 08} (2014) 115},
  \href{http://arxiv.org/abs/1405.6213}{{\ttfamily arXiv:1405.6213 [hep-th]}}.

\bibitem{Merbis:2014vja}
W.~Merbis, ``{Chern-Simons-like Theories of Gravity},''
  \href{http://arxiv.org/abs/1411.6888}{{\ttfamily arXiv:1411.6888 [hep-th]}}.

\bibitem{Li:2008dq}
W.~Li, W.~Song, and A.~Strominger, ``{Chiral Gravity in Three Dimensions},''
  \href{http://dx.doi.org/10.1088/1126-6708/2008/04/082}{{\em JHEP} {\bfseries
  04} (2008) 082}, \href{http://arxiv.org/abs/0801.4566}{{\ttfamily
  arXiv:0801.4566 [hep-th]}}.

\bibitem{Bergshoeff:2014pca}
E.~Bergshoeff, O.~Hohm, W.~Merbis, A.~J. Routh, and P.~K. Townsend, ``{Minimal
  Massive 3D Gravity},''
  \href{http://dx.doi.org/10.1088/0264-9381/31/14/145008}{{\em Class. Quant.
  Grav.} {\bfseries 31} (2014) 145008},
  \href{http://arxiv.org/abs/1404.2867}{{\ttfamily arXiv:1404.2867 [hep-th]}}.

\bibitem{Bergshoeff:2015zga}
E.~Bergshoeff, W.~Merbis, A.~J. Routh, and P.~K. Townsend, ``{The Third Way to
  3D Gravity},'' \href{http://dx.doi.org/10.1142/S0218271815440150}{{\em Int.
  J. Mod. Phys. D} {\bfseries 24} no.~12, (2015) 1544015},
  \href{http://arxiv.org/abs/1506.05949}{{\ttfamily arXiv:1506.05949 [gr-qc]}}.

\bibitem{Ozkan:2018cxj}
M.~\"Ozkan, Y.~Pang, and P.~K. Townsend, ``{Exotic Massive 3D Gravity},''
  \href{http://dx.doi.org/10.1007/JHEP08(2018)035}{{\em JHEP} {\bfseries 08}
  (2018) 035}, \href{http://arxiv.org/abs/1806.04179}{{\ttfamily
  arXiv:1806.04179 [hep-th]}}.

\bibitem{Alkac:2018eck}
G.~Alka\c{c}, M.~Tek, and B.~Tekin, ``{Bachian Gravity in Three Dimensions},''
  \href{http://dx.doi.org/10.1103/PhysRevD.98.104021}{{\em Phys. Rev. D}
  {\bfseries 98} no.~10, (2018) 104021},
  \href{http://arxiv.org/abs/1810.03504}{{\ttfamily arXiv:1810.03504
  [hep-th]}}.

\bibitem{Afshar:2019npk}
H.~R. Afshar and N.~S. Deger, ``{Exotic massive 3D gravities from
  truncation},'' \href{http://dx.doi.org/10.1007/JHEP11(2019)145}{{\em JHEP}
  {\bfseries 11} (2019) 145}, \href{http://arxiv.org/abs/1909.06305}{{\ttfamily
  arXiv:1909.06305 [hep-th]}}.

\bibitem{Bergshoeff:2013xma}
E.~A. Bergshoeff, S.~de~Haan, O.~Hohm, W.~Merbis, and P.~K. Townsend,
  ``{Zwei-Dreibein Gravity: A Two-Frame-Field Model of 3D Massive Gravity},''
  \href{http://dx.doi.org/10.1103/PhysRevLett.111.111102}{{\em Phys. Rev.
  Lett.} {\bfseries 111} no.~11, (2013) 111102},
  \href{http://arxiv.org/abs/1307.2774}{{\ttfamily arXiv:1307.2774 [hep-th]}}.
  [Erratum: Phys.Rev.Lett. 111, 259902 (2013)].

\bibitem{Afshar:2014dta}
H.~R. Afshar, E.~A. Bergshoeff, and W.~Merbis, ``{Interacting spin-2 fields in
  three dimensions},'' \href{http://dx.doi.org/10.1007/JHEP01(2015)040}{{\em
  JHEP} {\bfseries 01} (2015) 040},
  \href{http://arxiv.org/abs/1410.6164}{{\ttfamily arXiv:1410.6164 [hep-th]}}.

\bibitem{Ozkan:2019iga}
M.~Ozkan, Y.~Pang, and U.~Zorba, ``{Unitary Extension of Exotic Massive 3D
  Gravity from Bigravity},''
  \href{http://dx.doi.org/10.1103/PhysRevLett.123.031303}{{\em Phys. Rev.
  Lett.} {\bfseries 123} no.~3, (2019) 031303},
  \href{http://arxiv.org/abs/1905.00438}{{\ttfamily arXiv:1905.00438
  [hep-th]}}.

\bibitem{Bergshoeff:2021tbz}
E.~A. Bergshoeff, M.~Ozkan, and M.~S. Zog, ``{The holographic c-theorem and
  infinite-dimensional Lie algebras},''
  \href{http://dx.doi.org/10.1007/JHEP01(2022)010}{{\em JHEP} {\bfseries 01}
  (2022) 010}, \href{http://arxiv.org/abs/2110.09542}{{\ttfamily
  arXiv:2110.09542 [hep-th]}}.

\bibitem{Deger:2022gim}
N.~S. Deger, M.~Geiller, J.~Rosseel, and H.~Samtleben, ``{Minimal Massive
  Supergravity},'' \href{http://dx.doi.org/10.1103/PhysRevLett.129.171601}{{\em
  Phys. Rev. Lett.} {\bfseries 129} no.~17, (2022) 171601},
  \href{http://arxiv.org/abs/2206.00675}{{\ttfamily arXiv:2206.00675
  [hep-th]}}.

\bibitem{Bergshoeff:2019rdb}
E.~A. Bergshoeff, W.~Merbis, and P.~K. Townsend, ``{On asymptotic charges in 3D
  gravity},'' \href{http://dx.doi.org/10.1088/1361-6382/ab5ea5}{{\em Class.
  Quant. Grav.} {\bfseries 37} no.~3, (2020) 035003},
  \href{http://arxiv.org/abs/1909.11743}{{\ttfamily arXiv:1909.11743
  [hep-th]}}.

\bibitem{Ekiz:2022wbi}
E.~Ekiz, O.~Kasikci, M.~Ozkan, C.~B. Senisik, and U.~Zorba, ``{Non-relativistic
  and ultra-relativistic scaling limits of multimetric gravity},''
  \href{http://dx.doi.org/10.1007/JHEP10(2022)151}{{\em JHEP} {\bfseries 10}
  (2022) 151}, \href{http://arxiv.org/abs/2207.07882}{{\ttfamily
  arXiv:2207.07882 [hep-th]}}.

\bibitem{Deger:2023eah}
N.~S. Deger, M.~Geiller, J.~Rosseel, and H.~Samtleben, ``{Minimal massive
  supergravity and new theories of massive gravity},''
  \href{http://dx.doi.org/10.1103/PhysRevD.109.086014}{{\em Phys. Rev. D}
  {\bfseries 109} no.~8, (2024) 086014},
  \href{http://arxiv.org/abs/2312.12387}{{\ttfamily arXiv:2312.12387
  [hep-th]}}.

\end{thebibliography}\endgroup
		
	\end{document}